\newcommand{\nn}{\nonumber}

\newcommand{\be}{\begin{eqnarray}}
\newcommand{\ee}{\end{eqnarray}}

\documentclass[allclo]{FBSart}
\usepackage{amsfonts}
\usepackage{amssymb}
\title{Fermions in odd space-time dimensions: back to basics}
\author{A. Bashir  \thanks{\textit{E-mail address:} 
adnan@ifm.umich.mx} \instnr{1}, Ma. de Jes\'us Anguiano Galicia 
 \thanks{\textit{E-mail address:} 
marichuy@ifm.umich.mx}
\instnr{1,2}}
\instlist{Instituto de F{\'\i}sica y Matem\'aticas,
Universidad Michoacana de San Nicol\'as de Hidalgo, Apartado Postal
2-82, Morelia, Michoac\'an 58040, M\'exico
\and
Escuela de Ciencias F\'isico Matem\'aticas, 
Universidad Michoacana de San Nicol\'as de Hidalgo, Apartado Postal
2-82, Morelia, Michoac\'an 58040, M\'exico 
}

\runningauthor{A. Bashir, Ma. de Jes\'us Anguiano Galicia}
\runningtitle{Fermions in odd space-time dimensions: back to basics}
\begin{document}

\maketitle
\begin{abstract} 
It is a well known feature of odd space-time dimensions $d$ that there exist
two inequivalent fundamental representations $A$ and $B$ of the Dirac
gamma matrices. Moreover, the parity transformation swaps the fermion fields
living in $A$ and $B$. As a consequence, a parity invariant Lagrangian 
can only be constructed by incorporating both the representations.
Based upon these ideas and contrary to long held belief, we show that in
addition to a discrete exchange symmetry for the massless case, we can also
define chiral symmetry provided the Lagrangian contains fields corresponding 
to both the inequivalent representations. We also study the transformation
properties of the corresponding chiral currents under parity and charge 
conjugation operations. We work explicitly in $2+1$
dimensions and later show how some of these ideas generalize to an arbitrary
number of odd dimensions.
\end{abstract}


\maketitle

\section{Introduction}

There has been a growing interest in the last few years in theories with extra 
dimensions which have the capability of reducing the string mass scale 
$M_s$ several orders of magnitudes lower than the Planck mass scale of
$1.9 \times 10^{16}$ TeV, see e.g.,~\cite{extra}. 
In extreme scenarios,~\cite{extreme}, it can be as low as of the
order of TeV. In such cases, we can nip the gauge hierarchy problem in the bud.
In view of this exciting possibility, it is timely to revise various
properties of field theories in higher dimensions. Theories in even 
space-time dimensions are rather similar to the ones in 4-dimensions.
However, odd dimensions can have striking differences,~\cite{earlier,Shimizu}. 
In this
paper, we study the problem of chiral symmetry in odd space-time 
dimensions. We present a detailed discussion for the case of $d=3$
and then extend some of the ideas in a straightforward fashion to arbitrarily
high number of odd $d$. In the context of dynamical chiral symmetry breaking,
quantum electrodynamics in a plane (QED3) is interesting in its own 
right and this interest has been revived over the
past several years specially because the lattice results and the ones
obtained from the Shwinger-Dyson equation (SDE) studies,~\cite{differences},
have yet to
arrive at a final consensus. An interesting and somewhat uncomfortable 
observation is that
in the fundamental $2 \times 2$ representation of the gamma matrices,
chiral symmetry cannot even be defined owing to the fact there does not
exist any matrix which would anti-commute with all the three Pauli
matrices,~\cite{Pisarski,Appel}. This is in fact a feature of all 
odd dimensions $d$,~\cite{Shimizu}. It is for this reason
that the dynamical chiral symmetry breaking in a plane is generally 
studied in the $4 \times 4$ representation of the gammas, 
e.g.,~\cite{Pisarski,Appel,Roberts}.
There, we have two matrices anti-commuting with the gammas and
thus the same number of chiral transformations can be defined. 
With an even (2N) number
of 2-component fermions, one can combine two fermions to form two types of
mass terms, one is the usual chiral symmetry violating 
and the other the chiral symmetry preserving term. However, requiring parity 
conservation filters out the latter and we end up having a Lagrangian
similar in structure to the one in QED4. Although it is already known that
the four component spinor is in fact composed of two two-component
spinors, e.g., see the pioneering papers by Pisarski and 
Appelquist~\cite{Pisarski,Appel}, we do not find in literature explicit 
definition of chiral
transformations in the context of  two-component spinors alone without
any reference to the four-component spinors. 

   In the present paper, we take up this problem of whether chiral
symmetry can be defined in the fundamental representation of the gamma
matrices or not. For QED3, this is the $2 \times 2$ representation.
Naively, we can concentrate on any one set of gamma matrices specially 
because the corresponding spinors satisfy the well known completeness
relation. However, parity conservation, transformation properties of the 
fields under charge conjugation operation (in a certain set of 
odd dimensions)~\cite{Shimizu}, and chiral current 
conservation (for massless fermions) require taking into consideration the two
inequivalent representations.

   The work is organized as follows~: We start by considering a Lagrangian 
containing field corresponding to only one representation. We take up the 
parity conserving Lagrangian,~\cite{Shimizu}, and show
that the massless theory has more symmetries than the massive one,
a discrete exchange symmetry and two continuous symmetries. We can identify
the continuous symmetries with the chiral symmetries after calculating the
corresponding conserved currents and noting that they are the same as the
conventional chiral currents obtained by working in the $4 \times 4$ 
representation of the gamma matrices. We go on to show that in an obvious way, 
the same chiral transformations can be defined for any number of odd
dimensions in the fundamental representation of the gamma matrices.
We also study the transformation properties of the chiral currents under the
operation of parity and charge conjugation and make a comparison with results
known earlier in the context of 4-dimensional representation of 
QED3,~\cite{Burden}. We then present our conclusions.

\section{Symmetries}\label{II}

\noindent

Starting from the Lagrangian ${\cal L}=\bar{\psi} 
(i \hbar \not \! \partial - mc) \psi$, we can choose the following 
representation of the gamma matrices in a plane~: 
$\gamma^{0}=\sigma_3,\gamma^{1}=i\sigma_1$, and $\gamma^{2}=i\sigma_2$
(the $A$ representation hereafter), where 
$\sigma_i$ are the Pauli matrices. We then readily obtain the solutions of the
free Dirac equation~: 
\begin{eqnarray}
\nonumber
&& \hspace{-5mm} \psi_A^P(x)=
\left(\begin{array}{c} 1 \\  \frac{c(p_{y}-ip_{x})}{E+mc^{2}}
\end{array}\right)\,  e^{-\frac{i}{\hbar}x\cdot p} \equiv {u_A}(p)
e^{-\frac{i}{\hbar} x\cdot p} \;,   \\ \nonumber \\ 
&& \hspace{-5mm}
\psi_A^N(x)=
\left(\begin{array}{c} \frac{c(p_{y}+ip_{x})}{E+mc^{2}}  \\  1
\end{array}\right) \,  e^{\frac{i}{\hbar}x\cdot p} \equiv
{v}_A(p)
e^{\frac{i}{\hbar}x\cdot p} \;.  \label{rep1sol}
\end{eqnarray}
Choosing the normalization of the spinors to be such that there are
$2 E$ number of particles per unit volume, we have $
u_A \bar{u}_A = \not \!p c+mc^2$ and
$v_A \bar{v}_A =   \not \!p c-mc^2$.
This is to say that the completeness relations are not hampered by
the fact there is just one particle spinor and one anti-particle spinor. 
The above relation  permits us to define the projection operators 
$\Lambda_{\pm}= ({\pm \not \!p + mc})/{2mc}$ which project out the particle
and ant-particle spinors respectively. Therefore, everything is
apparently in order. However, there are reasons to believe 
that the above Lagrangian fails to incorporate various symmetries and
their consequences~:

\begin{itemize}

\item  {\bf Parity Invariance:} 
There are two independent solutions, one corresponding 
to a particle $(P)$ and the other to an anti-particle $(N)$. 
In a plane, there is just
one orbital angular momentum which we can define as
$L=r_x p_y - r_y p_x$. It does not commute with the Hamiltonian
$H=\gamma^0 (\vec{\gamma} \cdot \vec{p}+m c^2)$. However, if we define
the spin operator as $\Sigma= ({\hbar}/{2}) \gamma^0$, the total
angular momentum $J=L+\Sigma$ is a conserved quantity. It is easy to see 
that for the particle at rest, i.e., for $p_x=p_y=0$, $u_A$ and $v_A$
are eigenfunctions of $\Sigma$ with eigenvalues ${\hbar}/{2}$ and
-${\hbar}/{2}$ respectively. It implies a natural interpretation of the 
solution $u_A$ as that of a particle with spin say clockwise and of 
$v_A$ as that of an anti-particle with spin anti-clockwise. 
In a plane, just like any other odd dimensions, parity operation
is defined by reversing the signs of all but one coordinate. Let us 
suppose that under parity transformation, $r_x \rightarrow -r_x$ and
$r_y \rightarrow r_y$. Consequently, spin, being an angular momentum,
changes sign. Therefore, particle with clockwise spin
and anti-clockwise spin are related through the parity transformation. 
But one of these is not a solution of the Dirac equation. The Lagrangian 
and thus the particle spectrum are not parity invariant.

\item
{\bf Chiral Symmetry:}
In odd dimensions, all the anti-commutating matrices in the fundamental
representation are consumed by the Dirac gamma matrices. We are left with no
extra $\gamma_5$ which will anti-commute with all the gamma matrices.
Therefore, chiral transformations (and chiral symmetry) cannot be defined. 
Apparently, the massless Lagrangian has no more symmetry than the massive one.

\end{itemize}

 If we wish to incorporate these symmetries within the framework of
two-component spinors, we can do this thanks to the well-known fact that 
for odd $d$, there exist two inequivalent 
representations,~\cite{Shimizu}. In the planar case, we can choose 
$B$ to be $\gamma^{0}=\sigma_3,\gamma^{1}=i\sigma_1,\gamma^{2 \prime}
=- \gamma^2= -i\sigma_2$
We transform the corresponding solutions $\phi_B^P$ and $\phi_B^N$ of the 
Dirac equation to $\psi_B^P$ and $\psi_B^N$ for obvious particle
identification as follows~:
\begin{eqnarray}
\nonumber 
&& \hspace{-6mm} \psi_B^P(x)=i \gamma^2 \phi_B^P=  
\left(\begin{array}{c} \frac{c(p_{y}+ip_{x})}{E+mc^{2}}
\\   
 1 \end{array}\right)  e^{-\frac{i}{\hbar} p \cdot x}  = u_B(p) 
 e^{-\frac{i}{\hbar} p \cdot x}   \;, \\  \nonumber \\
&& \hspace{-6mm} \psi_B^N(x)  = i \gamma^2 \phi_B^N = \left(\begin{array}{c}
 1  \\ \frac{c(p_{y}-ip_{x})}{E+mc^{2}}
\end{array}\right)  e^{\frac{i}{\hbar} p \cdot x} 
= v_B(p)    e^{\frac{i}{\hbar} p \cdot x} \;. \label{rep2sol} \\ \nn
\end{eqnarray}
Looking at the stationary case, $p_x=p_y=0$, by applying the spin operator, 
we can see that $\psi_A^P$ and $\psi_B^P$ correspond to particles with
opposite spins. Similarly, $\psi_A^N$ and $\psi_B^N$ correspond to
anti-particles with opposite spins. The parity 
transformation ${\cal P}$~:
$(\psi_A)^{\cal P} = -i \gamma^1 \psi_B \; {\rm e}^{i \phi_1}$ 
and $(\psi_B)^{\cal P} = -i 
\gamma^1 \psi_A  \; {\rm e}^{i \phi_2}$ swaps the
spinors in the two inequivalent representations. It converts particle 
of one spin to the particle of opposite spin, and the same is true for
the anti-particle. The Lagrangian which takes into account both the 
representations is,~\cite{Shimizu}~: 
\begin{eqnarray}
{\cal L} &=& \bar{\psi_A} (i \hbar \not \! \partial - mc) \psi_A +
\bar{\psi_{B}} (i \hbar \not \!  \partial + mc) \psi_{B} \;.
\label{laginv}
\end{eqnarray}
It is parity invariant,~\cite{Shimizu}. The transformation properties 
of the fields under the 
charge conjugation operation ${\cal C}$ are\footnote{Note that the charge
conjugation operation does not mix the fields corresponding to the 
inequivalent representations for $d=3,7,11, \cdots$ but does so for
$d=1,5,9, \cdots$,~\cite{Shimizu}.}~:
$(\psi_A)^{\cal C} =  \gamma^2 (\bar{\psi}_A)^T \; {\rm e}^{i \psi_1}$ 
and $(\psi_B)^{\cal C} =   \gamma^2 
 (\bar{\psi}_B)^T   \; {\rm e}^{i \psi_2}$.
Most importantly, we shall show in 
the next section that this Lagrangian also permits us to define chiral 
symmetry.

\section{Chiral Symmetry}

We now pose ourselves the question whether the Lagrangian of 
Eq.~(\ref{laginv}) has any more symmetries for the massless case.
An immediate look reveals there is an exchange symmetry 
$\psi_A \leftrightarrow \psi_B$ which leaves the massless Lagrangian 
invariant. As it is only a discrete symmetry, it cannot be considered
a serious candidate for the chiral symmetry. However, one can
define following sets of simultaneous continuous transformations~: \newline
\noindent
{\em Set 1}
\begin{eqnarray}
\nonumber
\psi_A \; \rightarrow \; \psi_{A}^{\prime}
&=& \psi_{A} \; + \; \alpha \psi_{B} \;,
\\
\psi_{B}  \;\rightarrow \; \psi_{B}^{\prime} &=&
\psi_{B} \; - \; \alpha\psi_{A}\;.   \label{chiral1}
\end{eqnarray}
{\em Set 2}
\begin{eqnarray}
\nonumber
\psi_A \; \rightarrow \; \psi_{A}^{\prime}
&=& \psi_{A} \; + \; i \alpha \psi_{B} \;,
\\
\psi_{B}  \;\rightarrow \; \psi_{B}^{\prime} &=&
\psi_{B} \; + \; i \alpha\psi_{A}\;.   \label{chiral2}
\end{eqnarray}
Correspondingly, the Lagrangian~(\ref{laginv}) transforms in 
the following manner respectively~:
\begin{eqnarray}
{\cal L^{'}}_1 &=& {\cal L} \; - \; 2mc \alpha (\bar{\psi}_{A} \psi_{B}
\; + \; \bar{\psi}_{B} \psi_{A}) \;, \\
{\cal L^{'}}_2 &=& {\cal L} \; - \; 2 i mc \alpha (\bar{\psi}_{A} \psi_{B}
\; - \; \bar{\psi}_{B} \psi_{A}) \;,
\end{eqnarray}
Therefore, we conclude that under the continuous 
transformations~(\ref{chiral1},\ref{chiral2}), the massless
Lagrangian is invariant and the corresponding conserved currents are~:
\begin{eqnarray}
j^{\mu}_1\; &=& \; c \left( \bar{\psi}_{A}\;\gamma^{\mu}\;\psi_{B}
\; - \;
\bar{\psi}_{B}\;\gamma^{\mu}\;\psi_{A} \right)  \;,  
\label{chiralcurrent1}  \\
j^{\mu}_2 \; &=& \; c \left( \bar{\psi}_{A}\;\gamma^{\mu}\;\psi_{B}
\; + \;
\bar{\psi}_{B}\;\gamma^{\mu}\;\psi_{A} \right)  \;.
\label{chiralcurrent2}
\end{eqnarray}
To be able to identify continuous transformations~(\ref{chiral1}) as the
chiral transformation, we resort to address the issue of chiral symmetry 
in the often studied 4-dimensional representation of the gamma matrices
in the Lagrangian ${\cal L}=\bar{\psi} 
(i \hbar \not \! \partial - mc) \psi$~:
\begin{eqnarray} \gamma^0=
\left(\begin{array}{cc}
\sigma_3 & 0 \\ 0 & -\sigma_3
\end{array}\right)\;, \hspace{10mm}
\vec{\gamma}=
\left(\begin{array}{cc}
i \vec{\sigma} & 0  \\ 0  & -i \vec{\sigma}
\end{array}\right) \;,  \nn
\end{eqnarray}
where the vector $\vec{\sigma}$ has only two components, namely,
$\sigma_1$ and $\sigma_2$.
In the 4-dimensional 
representation, we have sufficient freedom to define two matrices which
anti-commute with the Dirac gamma matrices~:
\begin{eqnarray}
\nonumber
\gamma^{5} = i\left(\begin{array}{cc} 0 & I\\ -I & 0
\end{array}\right)  \; , \hspace{10mm}
\gamma^{3} = \left(\begin{array}{cc} 0 & I\\ I & 0
\end{array}\right)  \; ,
\nonumber
\label{e}
\end{eqnarray}
and hence we have two types of chiral transformations which yield the 
following chiral current in the massless limit~:
\begin{eqnarray}
j_5^{\mu} \equiv c \bar{\psi}\;\gamma^{\mu}\;\gamma^{5}\psi \;, 
 \hspace{10mm} j_3^{\mu} \equiv c \bar{\psi}\;\gamma^{\mu}\;\gamma^{3}\psi 
\;. \label{e2}
\end{eqnarray}
Interestingly, if we write out
\begin{eqnarray}
\psi = \left(\begin{array}{cc}
 \psi_A \\ \psi_B 
\end{array}\right) \;, \label{decomp}
\end{eqnarray}
the Lagrangian ${\cal L}=\bar{\psi} 
(i \hbar \not \hspace{-1mm} \! \partial - mc) \psi$ reduces to the one in 
Eq.~(\ref{laginv}). Therefore,
the components $\psi_A$ and $\psi_B$ themselves satisfy the Dirac equation
in the 2-dimensional representation of the gamma matrices justifying the
use of this notation. The 
chiral currents $j_5^{\mu}$ and  $j_3^{\mu}$ of Eq.~(\ref{e2}) can then
be written as~:
\begin{eqnarray}
\nonumber
j_5^{\mu} &=& c
\left(\bar{\psi}_{A}\;\; \bar{\psi}_{B}\right)\;\gamma^{\mu}\;
\left(\begin{array}{cc} 0 & 1\\ -1 & 0
\end{array}\right)\;
\left(\begin{array}{cc}
\psi_{A} \\ \psi_{B}
\end{array}\right)    \\
&=& c \left( \bar{\psi}_{A} \gamma^{\mu} \; \psi_{B}\; - \;
\bar{\psi}_{B} \gamma^{\mu}\;\psi_{A} \right)  \;, \label{e31} \\ \nn
\end{eqnarray}
and similarly
\begin{eqnarray}
j_3^{\mu} &=&  c \left( \bar{\psi}_{A} \gamma^{\mu} \; \psi_{B}\; + \;
\bar{\psi}_{B} \gamma^{\mu}\;\psi_{A} \right)  \;.  \label{e32}
\end{eqnarray}
Comparing Eqs.~(\ref{chiralcurrent1},\ref{e31}), we see that the currents
obtained from the conventional definition of chiral symmetry in the
4-dimensional representation and the ones we propose in the 2-dimensional 
representation, Eq.~(\ref{chiral1},\ref{chiral2}), are identical. {\em 
Therefore, Eq.~(\ref{chiral1},\ref{chiral2}) are indeed chiral 
transformations}. 

We know that the 4-component massless Dirac Lagrangian has a global U(2) 
symmetry generated by $\{1,\gamma_5,\gamma_3,i \gamma_3 \gamma_5 \equiv 
\gamma \} $. With the dynamical generation of fermion masses, this symmetry
is broken down to a $U(1) \times U(1)$ symmetry generated by 
$\{1, \gamma \}$. One may ask how does this latter symmetry manifests
itself in the $(\psi_A,\psi_B)$ space? We have already seen that the
symmetry corresponding to $\gamma_5$ and $\gamma_3$ mixes the fields 
$\psi_A$ and $\psi_B$. We do not expect the same for the residual 
symmetry and indeed such is the case. The following sets of transformations
correspond to the symmetry related to the generators
$1$ and $\gamma$, respectively~:  \newline
\noindent
{\em Set I}
\begin{eqnarray}
\nonumber
\psi_A \; \rightarrow \; \psi_{A}^{\prime}
&=& \psi_{A} \; + \; i \alpha \psi_{A} \;,
\\
\psi_{B}  \;\rightarrow \; \psi_{B}^{\prime} &=&
\psi_{B} \; + \; i \alpha\psi_{B}\;.   \label{unit}
\end{eqnarray}
{\em Set II}
\begin{eqnarray}
\nonumber
\psi_A \; \rightarrow \; \psi_{A}^{\prime}
&=& \psi_{A} \; + \; i \alpha \psi_{A} \;,
\\
\psi_{B}  \;\rightarrow \; \psi_{B}^{\prime} &=&
\psi_{B} \; - \; i \alpha\psi_{B}\;.   \label{gamma}
\end{eqnarray}
The corresponding conserved currents are
\begin{eqnarray}
j^{\mu}_I\; &=& \; c \left( \bar{\psi}_{A}\;\gamma^{\mu}\;\psi_{A}
\; + \;
\bar{\psi}_{B}\;\gamma^{\mu}\;\psi_{B} \right)  \;,  
\label{currentunit}  \\
j^{\mu}_{II} \; &=& \; c \left( \bar{\psi}_{A}\;\gamma^{\mu}\;\psi_{A}
\; - \;
\bar{\psi}_{B}\;\gamma^{\mu}\;\psi_{B} \right)  \;.
\label{currentgamma}
\end{eqnarray}
It is not hard to see that the line of reasoning similar to the one just
presented can be transported to higher odd dimensions 
smoothly. Let $
{\cal L}=\bar{\psi}  (i \hbar \not \! \partial - mc)  \psi  $
be the Lagrangian with doubly higher-dimensional representation of the
gamma matrices where chiral symmetry can be conventionally defined.
We denote the gamma matrices as $\gamma_{2^{(d+1)/2}}$. We can always choose
them to be
\begin{eqnarray} \gamma_{2^{(d+1)/2}}^{\mu}=
\left(\begin{array}{cc}
\gamma_{2^{(d-1)/2}}^{\mu} & 0 \\ \\ 0 & -\gamma_{2^{(d-1)/2}}^{\mu}
\end{array}\right)  \;,  \\ \nn
\end{eqnarray}
where $\gamma_{2^{(d-1)/2}}$ are the gamma matrices in the fundamental
representation. It is easy to show that
\begin{eqnarray*}
\{ \gamma_{2^{(d+1)/2}}^{\mu}, \gamma_{2^{(d+1)/2}}^{\nu} \} =
\{ \gamma_{2^{(d-1)/2}}^{\mu}, \gamma_{2^{(d-1)/2}}^{\nu} \} 
= 2 g^{\mu \nu} \;. 
\end{eqnarray*}
Therefore, again expressing $\psi$ as in Eq.~(\ref{decomp}), we get the
Lagrangian in Eq.~(\ref{laginv}). Note that in this generalized case,
$\psi_A$ and $\psi_B$ are ${2^{(d-1)/2}}$-dimensional, whereas, $\psi$ is
consequently ${2^{(d+1)/2}}$-dimensional. $\gamma_3$ and $\gamma_5$ retain
their definitions in the ${2^{(d-1)/2}}$ 
dimensions. Therefore, the transformations
of Eq.~(\ref{chiral1},\ref{chiral2}) are in fact chiral transformations in 
any number $d$ of odd dimensions.

\section{Transformation Properties of Currents}

In connection with the consequences of the advertised transformations
in the $(\psi_A,\psi_B)$ space to the positronium bound states, 
it may be interesting to look at the transformation properties of the
currents $j^{\mu}_I$, $j^{\mu}_{II}$, $j^{\mu}_1$ and 
$j^{\mu}_2$,~(\ref{currentunit},\ref{currentgamma}). One can easily show
that the currents $j^{\mu}_I$, $j^{\mu}_{II}$ transform
as follows~:
\begin{eqnarray*}
   \left( j^{\mu}_I \right)^{\cal P} = {\Lambda^{\mu}}_{\nu} 
j^{\mu}_I \;, \hspace{20 mm}   \left( j^{\mu}_{II} 
\right)^{\cal P} = - {\Lambda^{\mu}}_{\nu} 
j^{\mu}_{II} \;,
\end{eqnarray*}
where ${\Lambda^{\mu}}_{\nu} = {\rm diag}(1,-1,1)$. To obtain the reality of 
currents, we should define $j^{\mu}_1 \rightarrow i j^{\mu}_1$. On doing so 
and writing $\phi_1 - \phi_2\equiv \phi$, we have 
\begin{eqnarray}
\left(\begin{array}{c} j^{\mu}_1  \\ j^{\mu}_2
\end{array}\right)^{\cal P}   = \left(\begin{array}{cc} -{\rm cos} \phi &
-{\rm sin} \phi  \\  -{\rm sin} \phi & {\rm cos} \phi 
\end{array}\right)   \;  {\Lambda^{\mu}}_{\nu} 
\left(\begin{array}{c} j^{\nu}_1  \\ j^{\nu}_2 
\end{array}\right) \equiv R_p  \;  {\Lambda^{\mu}}_{\nu} 
\left(\begin{array}{c} j^{\nu}_1  \\ j^{\nu}_2 
\end{array}\right) \;.   \label{transPcurrent}
\end{eqnarray}
These are exactly the transformation properties given in~\cite{Burden} in the
context of 4-dimensional representation of the gamma matrices. 
In the spectrum of the ${\rm e}^+-{\rm e}^-$ bound sates for the dynamically 
broken chiral symmetry, as discussed in~\cite{Burden}, in addition to
scalar, pseudoscalar, vector and pseudovector ``mesons'', one also encounters
axi-scalar, axi-pseudoscalar, axi-vector and axi-pseudovector states in
accordance with the appearance of the parity matrix $R_p$. Here we see that
this matrix appears naturally in the $(\psi_A,\psi_B)$ space.
Defining $\psi_1 - \psi_2\equiv \psi$,  under the charge conjugation 
transformations
\begin{eqnarray}
\left(\begin{array}{c} j^{\mu}_1  \\ j^{\mu}_2
\end{array}\right)^{\cal C}   = \left(\begin{array}{cc} {\rm cos} \psi &
{\rm sin} \psi  \\  {\rm sin} \psi & - {\rm cos} \psi 
\end{array}\right)   \; 
\left(\begin{array}{c} j^{\mu}_1  \\ j^{\mu}_2 
\end{array}\right) \equiv R_c  \;  
\left(\begin{array}{c} j^{\mu}_1  \\ j^{\mu}_2 
\end{array}\right) \;.   \label{transCcurrent}
\end{eqnarray}
which also agree with the transformation properties of the corresponding
currents in the 4-dimensional representation 
tabulated in~\cite{Burden}. Hence, the transformation properties of the
currents studied in the 4-dimensional representation of the gamma matrices can 
be mapped nicely onto the $(\psi_A,\psi_B)$ space.

\section{Conclusions}\label{concl}

  We demonstrate that by taking into account both the inequivalent 
fundamental representations of the gamma matrices for odd number of 
space-time
dimensions, the resulting Lagrangian is not only parity invariant 
but also that  {\em we can write out chiral transformations within the
two-component description of the fermion spinors which mix the fields
belonging to the different inequivalent representations}. Interestingly,
these transformations reproduce the same chiral currents yielded by the 
conventional chiral symmetries defined in the doubly higher dimensional 
representation of the gamma matrices. In connection with the dynamically
broken chiral symmetry and the existence of the ${\rm e}^+-{\rm e}^-$
bound states in a plane, we also study the transformation properties of the 
currents under the operations of parity and charge conjugation
and find the expected correspondence with the studies carried out in the
4-dimensional representation of the gamma-matrices.

\section*{Acknowledgements}

Support for this work has been received in part by CIC under grant
number 4.10 and CONACyT under grant number 32395-E.

\end{document}